\documentclass[10pt]{article}
\usepackage[OE]{express}
\usepackage{braket}

\begin{document}
\title{Implementation and characterization of a controllable dephasing channel based on coupling polarization and spatial degrees of freedom of light}

\author{Daniel F. Urrego, \authormark{1} Juan-Rafael \'Alvarez, Omar Calder\'on-Losada, Ji\v{r}\'i Svozil\'ik, Mayerlin Nu\~nez and Alejandra Valencia~\authormark{2}}

\address{Laboratorio de \'Optica Cu\'antica, Universidad de los Andes, A.A. 4976, Bogot\'a, D.C., Colombia}

\email{\authormark{1}df.urrego1720@uniandes.edu.co} 
\email{\authormark{2}ac.valencia@uniandes.edu.co} 



\begin{abstract}
We present the experimental implementation and theoretical model of a controllable dephasing quantum channel using photonic systems. The channel is implemented by coupling the polarization and the spatial distribution of light, that play, in the perspective of Open Quantum Systems, the role of quantum system and environment, respectively. The capability of controlling our channel allows us to visualize its effects in a quantum system. Differently from standard dephasing channels, our channel presents an exotic behavior in the sense that the evolution of a state, from a pure to a mixed state, shows an oscillatory behavior if tracked in the Bloch sphere. Additionally, we report the evolution of the purity and perform a quantum process tomography to obtain the $\chi$ matrix associated to our channel.
\end{abstract}

\ocis{(200.0200) Optics in computing; (230.5440) Polarization-selective devices; (260.5430) Polarization; (270.2500) Fluctuations, relaxations, and noise; (270.5565) Quantum communications .} 


\bibliographystyle{osajnl}
\bibliography{Biblioteca}

\begin{thebibliography}{10}
\newcommand{\enquote}[1]{``#1''}

\bibitem{Nielsen2010}
M.~A. Nielsen and I.~L. Chuang, \emph{Quantum Computation and Quantum
  Information: 10th Anniversary Edition} (Cambridge University Press, New York,
  NY, USA, 2011), 10th ed.

\bibitem{Ricci2004}
M.~Ricci, F.~D. Martini, N.~J. Cerf, R.~Filip,
  J.~Fiur\'a\ifmmode~\check{s}\else \v{s}\fi{}ek, and C.~Macchiavello,
  \enquote{Experimental purification of single qubits,} Phys. Rev. Lett.
  \textbf{93}, 170501 (2004).

\bibitem{Lee2011}
J.-C. Lee, Y.-C. Jeong, Y.-S. Kim, and Y.-H. Kim, \enquote{Experimental
  demonstration of decoherence suppression via quantum measurement reversal,}
  Opt. Express \textbf{19}, 16309--16316 (2011).

\bibitem{Shaham2011}
A.~Shaham and H.~S. Eisenberg, \enquote{Realizing controllable depolarization
  in photonic quantum-information channels,} Phys. Rev. A \textbf{83}, 022303
  (2011).

\bibitem{Shaham2012}
A.~Shaham and H.~S. Eisenberg, \enquote{Realizing a variable isotropic
  depolarizer,} Opt. Lett. \textbf{37}, 2643--2645 (2012).

\bibitem{Jeong2013}
Y.-C. Jeong, J.-C. Lee, and Y.-H. Kim, \enquote{Experimental implementation of
  a fully controllable depolarizing quantum operation,} Phys. Rev. A
  \textbf{87}, 014301 (2013).

\bibitem{Fisher2012}
K.~A.~G. Fisher, R.~Prevedel, R.~Kaltenbaek, and K.~J. Resch, \enquote{Optimal
  linear optical implementation of a single-qubit damping channel,} New Journal
  of Physics \textbf{14}, 033016 (2012).

\bibitem{Salles2008}
A.~Salles, F.~de~Melo, M.~P. Almeida, M.~Hor-Meyll, S.~P. Walborn, P.~H.
  Souto~Ribeiro, and L.~Davidovich, \enquote{Experimental investigation of the
  dynamics of entanglement: Sudden death, complementarity, and continuous
  monitoring of the environment,} Phys. Rev. A \textbf{78}, 022322 (2008).

\bibitem{Kwiat2000}
P.~G. Kwiat, A.~J. Berglund, J.~B. Altepeter, and A.~G. White,
  \enquote{Experimental verification of decoherence-free subspaces,} Science
  \textbf{290}, 498--501 (2000).

\bibitem{Lemos2014}
G.~B. Lemos, J.~O. de~Almeida, S.~P. Walborn, P.~H.~S. Ribeiro, and
  M.~Hor-Meyll, \enquote{Characterization of a spatial light modulator as a
  polarization quantum channel,} Phys. Rev. A \textbf{89}, 042119 (2014).

\bibitem{Chuang1997}
I.~L. Chuang and M.~A. Nielsen, \enquote{Prescription for experimental
  determination of the dynamics of a quantum black box,} Journal of Modern
  Optics \textbf{44}, 2455--2467 (1997).

\bibitem{Fiurasek2001}
J.~Fiur\'a\ifmmode~\check{s}\else \v{s}\fi{}ek and Z.~c.~v. Hradil,
  \enquote{Maximum-likelihood estimation of quantum processes,} Phys. Rev. A
  \textbf{63}, 020101 (2001).

\bibitem{salazar15}
L.~J. Salazar-Serrano, A.~Valencia, and J.~P. Torres, \enquote{Tunable beam
  displacer,} Review of Scientific Instruments \textbf{86}, 033109 (2015).

\bibitem{James2001}
D.~F.~V. James, P.~G. Kwiat, W.~J. Munro, and A.~G. White, \enquote{Measurement
  of qubits,} Phys. Rev. A \textbf{64}, 052312 (2001).

\bibitem{Saleh2013}
B.~Saleh and M.~Teich, \emph{Fundamentals of Photonics}, Wiley Series in Pure
  and Applied Optics (Wiley, 2013).

\end{thebibliography}


\section{Introduction}

Many practical implementations of quantum-based applications must deal with the interaction of a quantum system and its environment, i.e. open quantum systems, Fig.~\ref{fig: dephasing}(a). This interaction can be described, in general, by a non-unitary operation associated with a quantum channel and it is responsible for the appearance of decoherence in the system. In the field of quantum information, for example, decoherence changes or destroys the information encoded in a quantum system, producing an incorrect transmission, processing or storage of it~\cite{Nielsen2010}.

Quantum channels can be classified depending on their effect in different input states. In the literature~\cite{Nielsen2010}, it is possible to find some specific examples: A depolarizing channel is one in which all pure states evolve towards the maximally mixed state. In an amplitude channel, all the pure states evolve towards a particular pure state that remains unaffected. A dephasing channel leaves two states, which are the opposite poles of the Bloch sphere, as decoherence-free states, while the rest of the pure states suffer decoherence.

Controllable channels make possible to visualize the effects of decoherence in a quantum system and to study the prevention of the appearance of such decoherence~\cite{Ricci2004, Lee2011}. Different types of channels that can be controlled have been performed using photonic systems. For example, depolarizing channels have been built using an arrangement of birefringent crystals ~\cite{Shaham2011, Shaham2012} and via interferometric effects~\cite{Jeong2013}. Amplitude channels have been performed in Ref.~\cite{Fisher2012, Salles2008}, and dephasing channels have been implemented using birefringent materials~\cite{Kwiat2000}, a spatial light modulator (SLM)~\cite{Lemos2014} or interferometric effects~\cite{Salles2008}.

In this paper, we report the experimental implementation and theoretical model of a controllable dephasing channel. Unlike a standard quantum dephasing channel, our device induces decoherence in which a pure state does not evolve linearly towards a mixed state, when seen in the Bloch sphere, but follows a spiral behavior. In our experiment, the polarization degree of freedom of light acts as the quantum system, and the continuous variable, corresponding to the spatial degree of freedom of light, represents the environment. The characterization of our channel is done by using the Bloch sphere representation and by measuring the purity of the output state. Additionally, using quantum process tomography~\cite{Chuang1997} (QPT) with a maximum-likelihood-estimation (MLE) algorithm~\cite{Fiurasek2001}, we report the $\chi$ matrix associated to our quantum channel.

\section{Theoretical Background}

\begin{figure}[htbp]
\centering
\includegraphics[scale = 0.84 ]{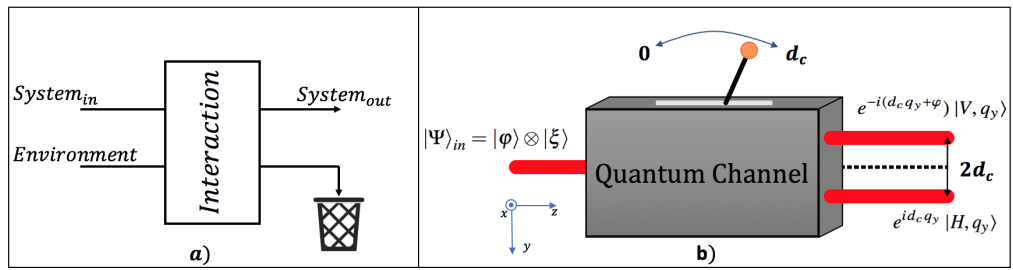}
\caption{(Color Online) (a) Model of an open quantum system. (b) An initial state passes through a quantum channel in which the original beam is separated into two parallel-propagating beams with orthogonal polarizations. The separation between the output beam centroids is $2d_c$ and occurs only in the $y$-direction.}
\label{fig: dephasing}
\end{figure}

In order to understand the working principle of our quantum channel, we start by presenting the theoretical model that describes its behavior. For this, consider a light beam that enters into the channel. Its polarization state can be represented by:
\begin{equation}
\ket{\varphi}=\alpha \ket{H} + \beta \ket{V},
\label{eq: inputpolo}
\end{equation}
where $\alpha$ and $\beta$ are probability amplitudes, satisfying $|\alpha|^2 + |\beta|^2=1$, and $\ket{H}$ and $\ket{V}$ are the horizontal and vertical polarization states, respectively. The channel couples such polarization state to the transverse momentum of light, whose state is given by
\begin{equation}
\ket{\xi}= \int d\vec{q} f(\vec{q}) \ket{\vec{q}},
\end{equation}
where $\vec{q}=\{q_x,q_y\}$ denotes the transverse momentum of light and the function $f(\vec{q})$ indicates the transverse momentum distribution.

As it has been mentioned, in our implementation the polarization and the transverse momentum of light represent the quantum system and the environment, respectively. The initial state, $\ket{\Psi}_{\text{in}}$ , that describes the beam entering into the quantum channel can be written as
\begin{equation}
\ket{\Psi}_{\text{in}}=\ket{\varphi} \otimes  \ket{\xi} = \alpha \int d\vec{q} f(\vec{q}) \ket{H,\vec{q}} + \beta \int d\vec{q} f(\vec{q}) \ket{V, \vec{q}}.
\label{eq: inputbeam}
\end{equation}

The channel is implemented by employing a polarizing tunable beam displacer (P-TBD)~\cite{salazar15}, a device that splits an incoming beam into two parallel propagating beams with orthogonal polarizations as depicted in Fig.~\ref{fig: dephasing}(b). The separation between the beams is restricted to the  $y$-direction and can be quantified by a tunable parameter, $2d_c$, that represents the distance between the centroids of the two output beams. Therefore, the channel can be represented by a unitary operation, $\hat{U}(d_c)$, that accomplishes the following transformations:
\begin{eqnarray}\label{eq: utransformation}
\hat{U}(d_c)\ket{H,q_y}&=& e^{i d_c q_y}\ket{H,q_y} \\ \nonumber
\hat{U}(d_c)\ket{V,q_y}&=& e^{-i( d_c q_y+\varphi)}\ket{V,q_y},
\end{eqnarray}
with $\varphi$ a generic phase difference that appears between the two output beams.

The density matrix of the output polarization state, $\hat{\rho}_{\text{out}}^{\text{pol}}$, can be written as

\begin{equation}
\hat{\rho}_{\text{out}}^{\text{pol}}=\text{Tr}_{\text{env}}\{\hat{U}(d_c)\ket{\Psi}_{\text{in}}\bra{\Psi}_{\text{in}}\hat{U}^{\dagger }(d_c) \},
\label{eq: rho_pol}
\end{equation}
where $\text{Tr}_{\text{env}}\{\bullet \}$ denotes the partial trace taken over the environment. Since our channel acts only in the $y$-direction, and $q_x$ and $q_y$ are independent, it is possible to write $f(\vec{q}) \propto f(q_y)$. In particular, for an input gaussian beam,
\begin{equation}
f(q_y) = N  e^{- w_y^2 \frac{(q_y-q_{0y})^2}{4}},
\label{uspatial}
\end{equation}
where $w_y$ is the beam's waist,  $q_{0y}$ is the center of its transverse momentum distribution, and $N$ is a normalization factor, such that $ \int |f(q_y)|^2 dq_y =1 $.

Plugging Eq.~(\ref{uspatial}) into Eq.~(\ref{eq: inputbeam}) and using  Eq.~(\ref{eq: rho_pol}), it is possible to write
\begin{equation}
\hat{\rho}_{\text{out}}^{\text{pol}} =\begin{pmatrix}
    |\alpha|^2 &  \alpha\beta^* e^{-\frac{2d_c^2}{w_y^2}}e^{i (2d_cq_{0y}+\varphi)} \\
        \alpha^*\beta e^{-\frac{2d_c^2}{w_y^2}}e^{-i (2d_cq_{0y}+\varphi)} & |\beta|^2 \\
        \end{pmatrix}.
\label{eq: rho_final}
\end{equation}
From this density matrix, it is clearly seen that the implemented channel changes the initial polarization state of the beam depending on the value of the parameter $d_c$. The off-diagonal terms contain the expression $e^{-\frac{2d_c^2}{w_y^2}}e^{i (2d_cq_{0y}+\varphi)}$ or its conjugate. The term $e^{-\frac{2d_c^2}{w_y^2}}$ is responsible for the decay of a pure state into a mixed state, revealing the decoherence of the system induced by a standard dephasing channel~\cite{Nielsen2010}. However, the presence of $e^{i (2d_cq_{0y}+\varphi)}$ indicates that our channel does not behave as a standard dephasing.

The characterization of the channel is performed by  obtaining the Stokes parameters, $\{S_0, S_1, S_2, S_3\}$, associated with a polarization state that passes through it. These parameters are represented in the Bloch sphere, and the corresponding purity can also be obtained. The Stokes parameters can be extracted from the polarization density matrix~\cite{James2001} given in Eq.~\ref{eq: rho_final},
\begin{equation}
S_0= |\alpha|^2   + |\beta|^2,
\label{eq: stokes_exp s0}
\end{equation}
\begin{equation}
S_1=|\alpha|^2   - |\beta|^2,
\label{eq: stokes_exp s1}
\end{equation}
\begin{equation}
S_2(d_c)=  2|\alpha| | \beta^* |e^{-\frac{2 d_c^2}{ w_y^2}} \cos(2q_{0y} d_c +\varphi+\Phi),
\label{eq: stokes_exp s2}
\end{equation}
\begin{equation}
S_3(d_c)= 2|\alpha| | \beta^* |e^{-\frac{2 d_c^2}{ w_y^2}} \sin(2q_{0y} d_c +\varphi+\Phi),
\label{eq: stokes_exp s3}
\end{equation}
where $\Phi=\Phi_\alpha - \Phi_\beta$ with $\alpha=|\alpha|e^{i\Phi_\alpha}$ and $\beta=|\beta|e^{i\Phi_\beta}$.

\begin{figure}[htbp]
\centering\includegraphics[scale=0.7]{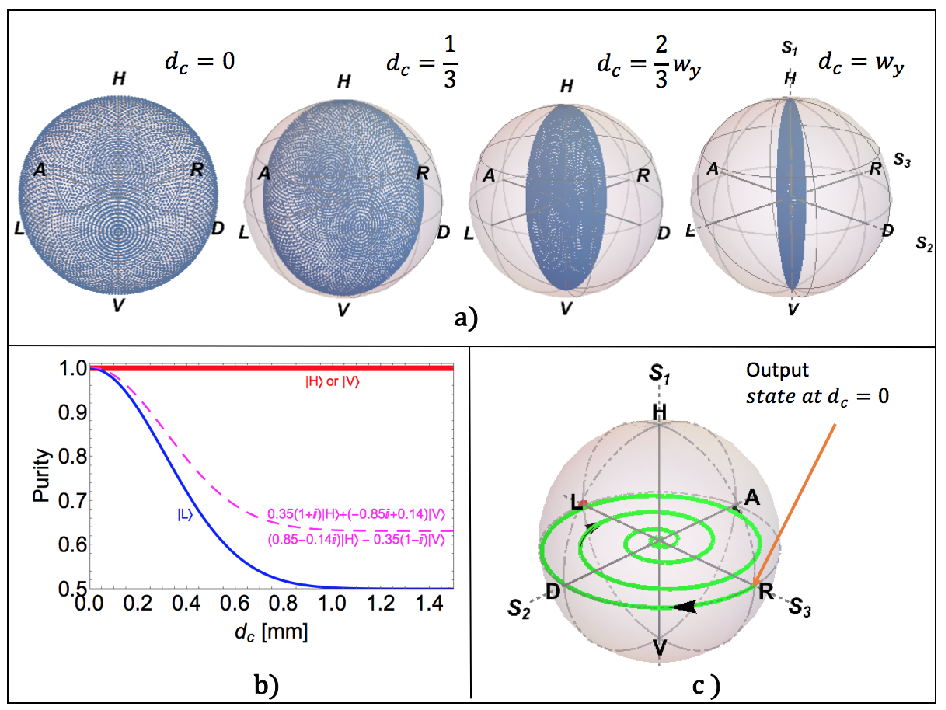}
\caption {(Color Online) (a) It shows, in the Bloch sphere, the predicted effect of our channel on various input pure states. The dark regions represent the output polarization states. The spheres correspond to different settings of the parameter $d_c$. (b) It shows the theoretical evolution of the purity  for five different input polarization states (Table~\ref{tab: data}) as the value of $d_c$ varies. (c) It shows the evolution of an input left circular polarization  state with the parameter $d_c$. }
\label{fig: teo}
\end{figure}

Figure~\ref{fig: teo} shows the effect of our channel on various input pure polarization states when $\varphi = \pi$ and they are coupled to an environment  characterized by a Gaussian distribution with $w_y=0.88$~mm and $q_{0y}=10.6~\text{mm}^{-1}$. The values of $\varphi$, $q_{0y}$ and $w_y$ are chosen to match the experimental parameters as will be described in sections~(\ref{sec: 3})-(\ref{sec: 4}). In Fig.~\ref{fig: teo}(a), the dark regions represent the output polarization states in the Bloch sphere. Each polarization state is defined by a set of coordinates given by the Stokes parameters $\{S_1,S_2(d_c),S_3(d_c)\}$. The depicted spheres correspond to different settings of the parameter $d_c$, specifically, $d_c=0$, $d_c=\frac{w_y}{3}$, $d_c=\frac{2w_y}{3}$ and $d_c=w_y $. For the case $d_c=0$, only a dark sphere is observed, since the channel only induces a rotation maintaining the output state pure and therefore on the surface of the Bloch sphere. In contrast, when $d_c$ increases, the dark region indicates that the poles of the sphere remain unchanged while the rest of the sphere shrinks towards the vertical axis, $S_1$. This behavior indicates that horizontal and vertical polarizations are decoherence free-states while any other input pure state becomes mixed. This particular way in which the Bloch sphere is shrinking is a signature of the fact that the proposed coupling corresponds to a dephasing channel.

Another way to identify the type of channel is by means of the evolution of the purity of the output polarization state, $P_{\text{out}}$. Using the Stokes parameters, $P_{\text{out}}$ can be written as $P_{\text{out}}=\text{Tr}\{(\hat{\rho}_{\text{out}}^{\text{pol}})^2\}=(S_0^2+S_1^2+S_2^2+S_3^2)/2$. Using Eq.~(\ref{eq: stokes_exp s0}), Eq.~(\ref{eq: stokes_exp s1}), Eq.~(\ref{eq: stokes_exp s2})  and  Eq.~(\ref{eq: stokes_exp s3}) the purity becomes
\begin{equation}
P_{\text{out}}(d_c)=\frac{1}{2}\left(1+\left(|\alpha|^2   - |\beta|^2\right)^2 + 4 |\alpha  \beta^* |^2 e^{-\frac{4 d_c^2}{w_y^2}} \right).
\label{eq: purity-2}
\end{equation}

The behaviour of Eq.~(\ref{eq: purity-2}) is depicted in Fig.~\ref{fig: teo}(b) for five different initial polarization states defined in Table~\ref{tab: data}. The wide solid line corresponds to horizontal and vertical polarizations, the dashed line refers to the polarization states defined by $\{\alpha=0.85-0.14i, \beta= -0.35 (1- i)\}$ and $\{\alpha=0.35(1+i), \beta=0.14-0.85i\}$, and the thin solid line corresponds to the purity of left circular polarization $\{\alpha=\frac{1}{\sqrt{2}}, \beta=\frac{-i}{\sqrt{2}}\}$. From these graphs, it is possible to see that if the input state has either horizontal or vertical polarization, the purity is one. This means that the channel does not affect the system and therefore there is no decoherence. On the other hand, for the other pure input states, the purity starts at one and monotonically decreases when $d_c$ increases reaching a value $P_{\text{out}}(d_c) \approx \frac{1}{2}(S_1^2+1)$ for  $d_c\gg w_y$. According to the value of $S_1$ reported in Table~\ref{tab: data}, it is possible to see that for a state evolving in the equator the purity tends to $0.5$, while for a state in the Tropic of Cancer or Tropic of Capricorn, the purity tends to $0.625$.

\begin{table}
\begin{center}
\textbf{\caption{Initial polarization states }\label{tab: data}}
\begin{tabular}{|c c|c|c|}
  \hline
  \textbf{$\alpha$} & \textbf{$\beta$} & \textbf{$S_1$} & \bf Name  \\ \hline
   1 & 0 & 1& North Pole  \\
  $0.85-0.14 i$ & $-0.35 (1-i)$ & 0.5   & Tropic of Cancer  \\
  $\frac{1}{\sqrt{2}}$ & $ \frac{-i}{\sqrt{2}}$ & 0 & Equator  \\
  $0.35(1+i)$ & $0.14-0.85i$ & -0.5 & Tropic of Capricorn \\
  0 & 1 & -1 & South Pole \\
  \hline
\end{tabular}
\end{center}
\end{table}

Figure~\ref{fig: teo}(c) shows the evolution of an input left circular polarization state (located in the equator plane) with the parameter $d_c$. It is seen that, as $d_c$ increases, the state tracks a spiral on the equatorial plane of the Bloch sphere. This spiral starts in the surface, for $d_c=0$, and ends up in the center. This type of evolution is not typical for a dephasing channel and gives an exotic characteristic to the implementation we report. The spiral evolution is due to the presence of oscillatory terms in the off-diagonal elements of $\hat{\rho}_{\text{out}}^{\text{pol}}$ (Eq.~\ref{eq: rho_final}). These terms are caused by the fact that our environment corresponds to a Gaussian  distribution  centered at $q_{0y}$. Physically, $q_{0y}$ is related to a small deviation of the incoming beam with respect to the $z$-direction~\cite{Saleh2013}, given by an angle $\theta=\frac{q_{0y} \lambda}{2 \pi}$ with $\lambda$ the wavelength of the light beam. A standard dephasing channel can be recovered by setting $q_{0y}=0$.

There are two interesting ideas to notice from Fig.~\ref{fig: teo}(c): first, at $d_c=0$ the output state has right circular polarization, even though the input state was initially prepared as left circular. This occurs because of the presence of the phase $\varphi=\pi$, that comes from the experimental implementation of the channel. Second, although Fig.~\ref{fig: teo}(c) is for an input left circular polarization state, any other input pure state will follow the spiral behavior in a plane parallel to the equator of the Bloch sphere.

\section{Experimental Implementation}\label{sec: 3}

\begin{figure}[htbp]
\centering\includegraphics[scale=0.82]{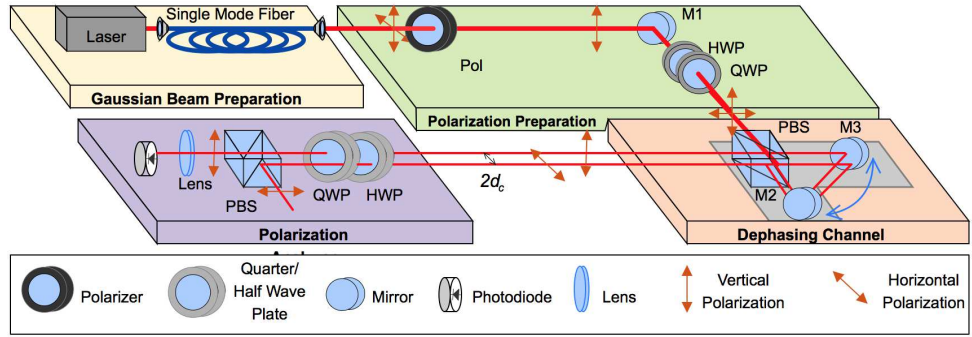}
\caption{(Color Online) Experimental setup composed of four steps. The first step is the preparation of a Gaussian spatial mode. The second step is the preparation of the polarization state. The third step is the experimental implementation of our dephasing channel. Finally, the fourth step is the polarization analyzer.}
\label{fig: setup}
\end{figure}

Figure~\ref{fig: setup} shows the setup for the implementation and characterization of our channel. Four steps can be clearly recognized. In the first step, an 808 nm-CWlaser (Thorlabs, CPS808) is coupled into a single mode fiber to obtain a Gaussian beam with a waist around $w_y=0.88$~mm and that can be considered collimated during the whole path of the experiment. In the second step, a polarizer is set to fix a vertical polarization; a quarter wave plate (QWP) and a half wave plate (HWP) are placed to obtain different polarizations that defined the input state by setting the values $\alpha$ and $\beta$ in Eq.~(\ref{eq: inputpolo}). The third step in our setup is the channel implemented by employing a polarizing tunable beam displacer (P-TBD)~\cite{salazar15}. It consists of a polarizing beam splitter (PBS) and two mirrors, $M2$ and $M3$, placed on an L-shaped platform that is mounted on a rotational stage. By rotating this platform, the separation $2d_c$ between the two emerging beams can be tuned as illustrated in Fig.~\ref{fig: dephasing}(b). It is relevant to notice that due to the working principle of the P-TBD, the two beams that come out from it suffer a different amount of reflections in the PBS, $M2$ and $M3$. This fact justifies the introduction of $\varphi$ in the theoretical model. The different amount of reflections results precisely in $\varphi=\pi$, explaining our choice of $\varphi$ for the graphs of the theoretical model in Figure~\ref{fig: teo}. In the fourth step, a polarization tomography analysis is implemented for different values of the separation $d_c$. This process is performed by sending the light through a HWP, a QWP and a PBS. The light transmitted by the PBS is focused, with a lens ($f=25.4$~mm), into a photodiode (Thorlabs FDS100) while the light coming from the reflecting output is neglected. The Stokes parameters are then recovered by performing intensity measurements in the photodiode for different settings of the HWP and QWP.

\section{Results and discussion}\label{sec: 4}

\begin{figure}[htbp]
\centering\includegraphics[scale=0.85]{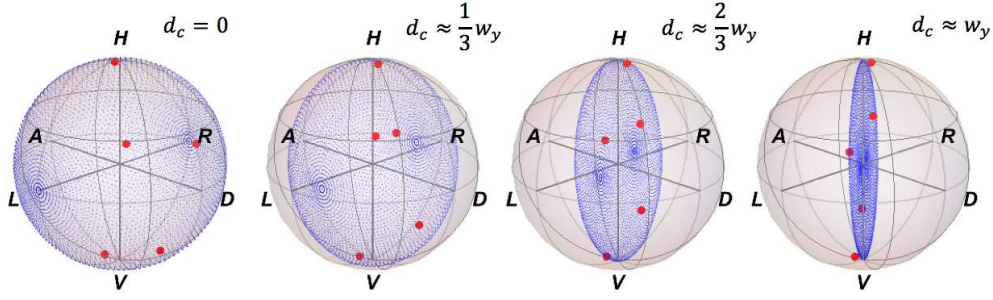}
\caption{(Color Online)  Bloch sphere representation of the output polarization states (red dots) measured  for the five initial input states defined in Table~\ref{tab: data}. The dark region is the same as the one reported in the theoretical section, Fig.~\ref{fig: teo}(a).}
\label{fig: resultsBloch-shrink}
\end{figure}

The characterization of our channel is done by measuring how the Bloch sphere shrinks for different values of the parameter $d_c$ when various input states are considered. The measurement of the Stokes parameters is done for the five different input polarization states defined in Table~\ref{tab: data}. Figure~\ref{fig: resultsBloch-shrink} shows as dots the experimental data corresponding to $\{S_1,S_2(d_c), S_3(d_c)\}$ for the values $d_c=0$, $d_c=\frac{w_y}{3}$, $d_c=\frac{2w_y}{3}$ and $d_c=w_y $. The dark region is the same as the one reported in the theoretical section, Fig.\ref{fig: teo}(a), and it is clearly seen that the experimental data are contained on it. From the way in which the shrinking of the Bloch sphere occurs according to Fig.~\ref{fig: resultsBloch-shrink}, it is possible to conclude that indeed the channel presented in this work induces dephasing.

\begin{figure}[htbp]
\centering\includegraphics[scale=0.85]{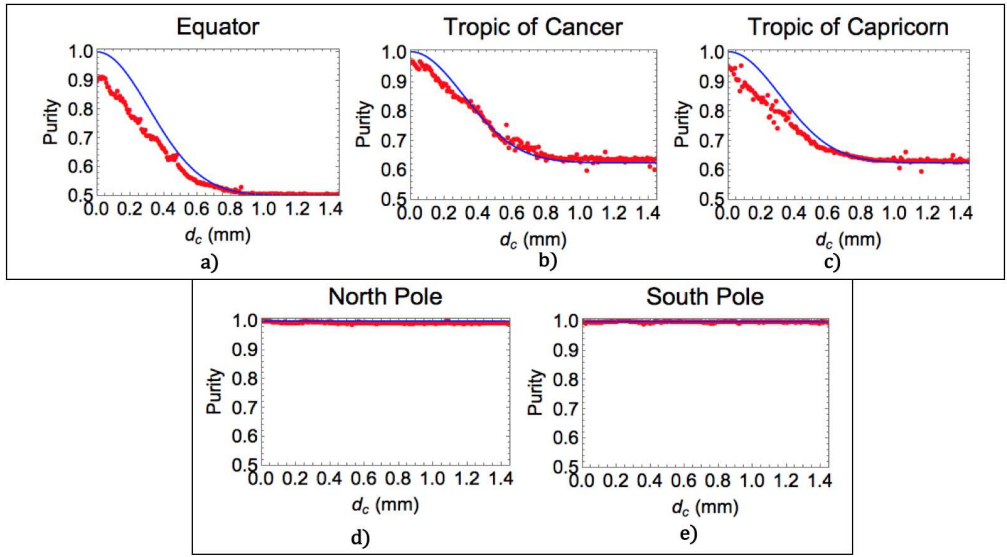}
\caption{(Color Online) The points are the experimental results for the purity as a function of the parameter $d_c$. The solid lines correspond to the theoretical prediction for the purity. Each graph shows the behavior of the initial input state defined in Table~\ref{tab: data}. }
\label{fig: purity}
\end{figure}

The experimental results for the evolution of the purity, $P_{\text{out}}(d_c)$, are shown in  Fig. \ref{fig: purity} for the five states of Table~\ref{tab: data}. The dots are experimental data corresponding to $\{S_1,S_2(d_c), S_3(d_c)\}$ when $d_c$ is scanned in the range $[0, 1.42]$~$\text{mm}$ in steps of $7.2$~$\mu \text{m}$. The solid lines correspond to the theoretical model  according to Eq.~(\ref{eq: purity-2}). As expected, the purity for the horizontal and vertical input states remain unchanged; for an input state in the Tropic of Cancer or the Tropic of Capricorn, $P_{\text{out}}(d_c)$ tends to $0.625$ while for left circular polarization,  $P_{\text{out}}(d_c)$ tends to $0.5$. The deviation between the experimental data and the theoretical model is due to  technical features associated with the optical elements and the uncertainty on the rotation angle of the L-shaped platform of the P-TBD.

\begin{figure}[htbp]
\centering\includegraphics[scale=0.85]{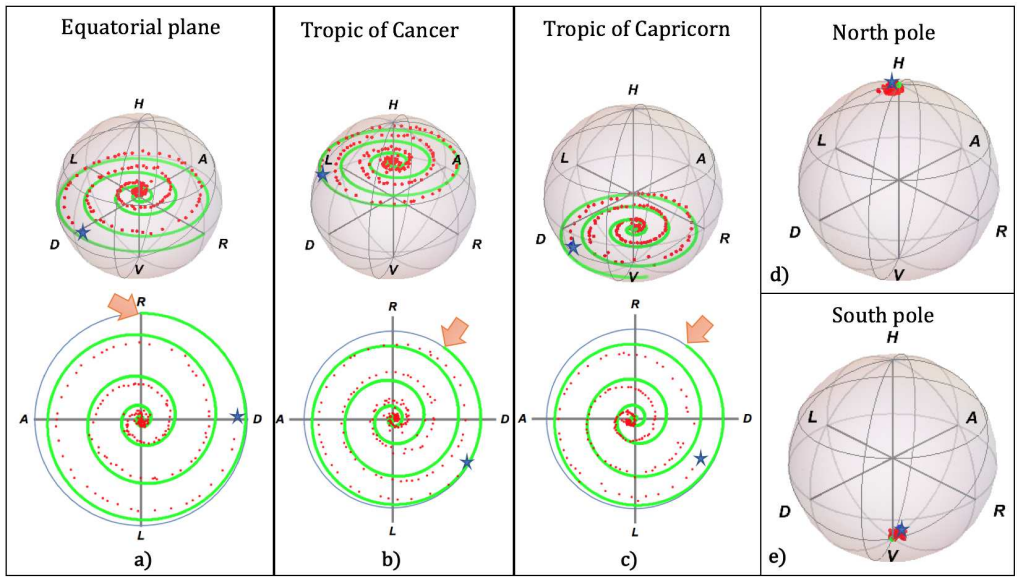}
\caption{(Color Online) In the Bloch sphere representation, the path followed by a state is tracked when the parameter $d_c$ is evolved. The dots are the experimental data and the solid lines are the theoretical model. The big blue stars are the initial input states defined in Table~\ref{tab: data}.  Graphs (a)-(c) shown the exotic spiral that the states follow. The graphs (d)-(e) show that horizontal and vertical input states do not change.}
\label{fig: resultsBloch}
\end{figure}

The exotic behavior of our channel, in which a state suffers decoherence by following a spiral in the Bloch sphere, is corroborated in our experiment by tracking the evolution of a left circular input polarization state when the parameter $d_c$ is varied. Figure~\ref{fig: resultsBloch}(a) depicts this evolution in the Bloch sphere and its projection on the equatorial plane. The values of $d_c$ are the same used in the measurement of the purity. The dots are experimental data, corresponding to the measurement of $\{S_1,S_2(d_c), S_3(d_c)\}$,  and the solid line is the theoretical model according to Eq.~(\ref{eq: stokes_exp s1}), Eq.~(\ref{eq: stokes_exp s2}) and Eq.~(\ref{eq: stokes_exp s3}) using $\varphi=\pi$ and $q_{0y}$  as a fitting parameter with a value of $q_{0y}^{\text{fit}}=10.6~\text{mm}^{-1}$. The star indicates the output state that is measured at $d_c=0$ and the big arrow indicates the theoretical output state at $d_c=0$. The discrepancy between these two points is due to the same experimental uncertainty that was mentioned when discussing the experimental data for the purity in Fig.~\ref{fig: purity}.

As complementary measurements, the evolution of the other four states of  Table~\ref{tab: data} was also tracked. The data and theoretical model for input states in the Tropic of Cancer and Tropic of Capricorn are shown in  Figs.~\ref{fig: resultsBloch}(b)-\ref{fig: resultsBloch}(c) as dots and solid lines, respectively. From these graphs, it is clearly seen that the pure input states enter inside the Bloch sphere following a spiral parallel to the equator with a latitude defined by $S_1$. The lower plots are the projection of the corresponding planes. Figs.~\ref{fig: resultsBloch}(d)-\ref{fig: resultsBloch}(e) reveal that, as expected for a dephasing channel, the horizontal and vertical input states remain in the poles.

\begin{figure}[htbp]
\centering\includegraphics[scale=0.6]{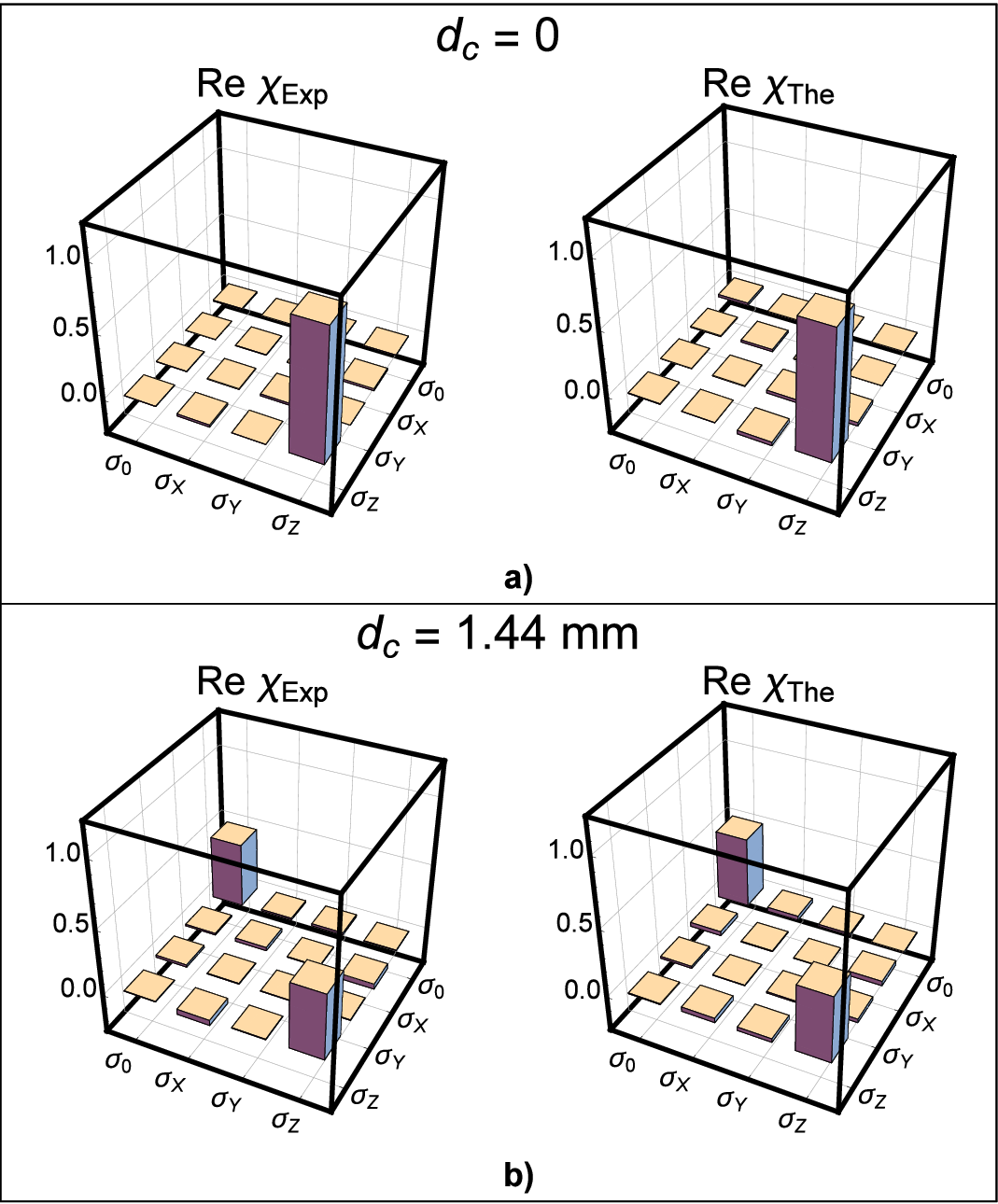}
\caption{(Color Online) The real parts of the experimental and the theoretical $\chi$ matrix are shown. (a) $\chi$ matrix for $d_c=0$. (b) $\chi$ matrix for $d_c = 1.44$~mm. The vanishing imaginary parts of $\chi$ are not shown.}
\label{fig: chi_exp_teo}
\end{figure}

In order to complete the characterization of the described channel, we implemented a quantum process tomography method \cite{Chuang1997,Fiurasek2001} to obtain its associated $\chi$ matrix. The complete positive map related to our channel is given by
 \begin{equation}
 \hat{\rho}_{\text{out}}^{\text{pol}}=\sum_{i,j} \chi_{ij} \hat{\sigma}_i \hat{\rho}_{in}^{\text{pol}} \hat{\sigma}^{\dagger}_j, 
 \label{eq:map}
 \end{equation} 
where $\hat{\rho}_{in}^{\text{pol}}= \ket{\varphi} \bra{\varphi}$, with $\ket{\varphi}$ defined in Eq.~(\ref{eq: inputpolo}), and  $\hat{\sigma}_i$ are the matrices that span the space of $\hat{\rho}_{in}^{\text{pol}}$ that correspond, in this case, to the identity ($\hat{\sigma}_0=\hat{I}_{2\times2}$) and the Pauli matrices ($\hat{\sigma}_X$, $\hat{\sigma}_Y$ and $\hat{\sigma}_Z$). Figure~\ref{fig: chi_exp_teo} shows the  real part of   the experimental and theoretical $\chi$ matrices, $\chi_{\text{Exp}}$ and $\chi_{\text{The}}$, for $d_c=0$ and $d_c = 1.44$~mm, corresponding to the initial and final values of $d_c$ used in our experiment. To obtain $\chi_{\text{Exp}}$, Eq.~(\ref{eq:map}) was solved, using  MLE with the experimental data. For $\chi_{\text{The}}$, MLE was used with $\hat{\rho}_{\text{out}}^{\text{pol}}$ given by Eq.~(\ref{eq: rho_final}) with $\alpha$ and $\beta$ shown in Table~\ref{tab: data}. From Fig.~\ref{fig: chi_exp_teo}(a), it is clearly observed that our channel does not induce decoherence when $d_c=0$ but generates only a rotation. This rotation is precisely the one that mentioned before as coming from  the working principle of the channel (P-TBD). For $d_c = 1.44$~mm, Fig.~\ref{fig: chi_exp_teo}(b) shows that the diagonal of the $\chi$ matrix has contributions from the identity and $\hat{\sigma}_Z$. This fact can be related with the shrinking of the Bloch sphere depicted in Fig.~\ref{fig: teo}(a), indicating that our channel indeed corresponds to a dephasing one.

\section{Conclusions}

In this paper, we have presented  the theoretical model and the experimental implementation of a controllable dephasing channel. The channel is implemented using photonics systems and considering polarization as a quantum system and the environment simulated by the transverse momentum distribution of light. The implemented channel has been performed with the help of a Polarizing Tunable Beam Displacer, which permits the control of the decoherence that the polarization system suffers as a function of the tunable parameter, $d_c$. In order to identify the type of quantum channel that our setup describes, we have observed its effects in the polarization by means of the Bloch sphere representation, we have measured the evolution of the purity and we have performed a QPT protocol to identify the $\chi$ matrix. When the channel's parameter, $d_c$, is tuned, the Bloch sphere and the purity evolve in a manner that corresponds to a standard dephasing channel. Interestingly, the implemented channel exhibits also an exotic characteristic in which the evolution toward mixed states is done by following a spiral path. This deviation from the standard is due to the characteristics of the environment that are considered for this implementation. The results reported in this paper can be useful for decoherence suppression implemented via quantum error correction protocols in quantum information applications.

\section*{Acknowledgments}

The author's acknowledge support from Facultad de Ciencias, Universidad de Los Andes. AV and MNP acknowledge support from FAPA project from Universidad de los Andes, Bogotá, Colombia.

\end{document}